\documentclass[twocolumn,floats,prl,aps,showpacs,superscriptaddress,citeautoscript,floatfix]{revtex4}

\usepackage[dvips]{graphicx}
\usepackage{hyperref}
\usepackage{dcolumn}

\newcommand{\icm}{\ensuremath{~\textrm{cm}^{-1}}}	

\newcolumntype{.}{D{.}{.}{-1}}

\begin{document}

\title{Infrared phonon dynamics of multiferroic BiFeO$_3$ single crystal}

\author{R.P.S.M. Lobo}
\email{lobo@espci.fr}
\affiliation{Laboratoire Photons et Mati\`ere (CNRS UPR 5), ESPCI, Universit\'e Pierre et Marie Curie, 10 rue Vauquelin, 75231 Paris Cedex 5, France}

\author{R.L. Moreira}
\affiliation{Departamento de F\'{\i}sica, ICEx, Universidade Federal de Minas Gerais, CP 702, Belo Horizonte, MG, 31270-901, Brazil}

\author{D. Lebeugle}
\author{D. Colson}
\affiliation{Service de Physique de l'\'Etat Condens\'e (CNRS URA 2464), DSM/DRECAM/SPEC, CEA Saclay 91191 Gif sur Yvette Cedex, France}

\date{\today}

\begin{abstract} 
We discuss the first infrared reflectivity measurement on a BiFeO$_3$ single crystal between 5 K and room temperature. The 9 predicted \textit{ab}-plane $E$ phonon modes are fully and unambiguously determined. The frequencies of the 4 $A_1$ \textit{c}-axis phonons are found. These results settle issues between theory and data on ceramics. Our findings show that the softening of the lowest frequency $E$ mode is responsible for the temperature dependence of the dielectric constant, indicating that the ferroelectric transition in BiFeO$_3$ is soft-mode driven.
\end{abstract}

\pacs{78.30.-j, 77.84.-s, 77.22.-d}

\maketitle


The interplay between magnetic and dielectric properties is an intriguing subject that was already discussed by Pierre Curie in 1894 \cite{Curie}. Such, magnetoelectric effects were subject of extensive studies during the 1960s \cite{Freeman, Smith, Moreau}. More recently, the ability of controlling the dielectric properties through a magnetic field and \textit{vice-versa} renewed the interest in these materials \cite{Huang, Katsufuji, Kimura}. Coupling between electrical and magnetic properties are particularly interesting in multiferroic materials, which present simultaneously two or more ferroic or antiferroic order parameters. The ferroelectric \cite{Smith} and antiferromagnetic properties of BiFeO$_3$ based systems, showing $T_\textrm{Curie} = 1090$ K \cite{Lebeugle2007} and $T_\textrm{N\'eel} = 673$ K \cite{Moreau}, have been known for a long time but the possibility of making devices operating at room temperature \cite{Zhao} gave a new burst to bismuth ferrite. The coexistence of ferroelectricity and antiferromagnetism at room temperature, makes BiFeO$_3$ one of the most interesting multiferroic materials. 

Optical spectroscopy is a very powerful tool to understand the driving mechanism of the ferroelectric transition and, eventually, its coupling to magnetic ordering \cite{Souchkov, Litvinchuk}. Nevertheless, only a few Raman \cite{Haumont, Fukumura, Cazayous} and infrared (IR) \cite{Kaczmarek, Kamba} studies on BiFeO$_3$ are known to date. Surprisingly none of the IR measurements were carried out on single crystals. There are two important open questions concerning the electrodynamics of BiFeO$_3$: (i) the phonon modes symmetry; and (ii) the role (or absence of thereof) of phonon softening in the ferroelectric transition. 

In its ferroelectric phase, BiFeO$_3$ belongs to the $R3c$ ($C_{3v}^6$) space group, which derives from the ideal (paraelectric) $Pm\overline{3}m$ ($O_h^1$) cubic perovskite group by a small distortion along the $\left(111\right)$ cubic directions. Its magnetic structure is a spiral-cycloidal incommensurate phase. Group theory analysis for the $R3c$ ferroelectric group predicts $4 A_1 \oplus 5 A_2 \oplus 9 E$ optical phonon modes \cite{Kamba}. The $A_1$ modes are infrared active along the \textit{c}-axis, the $E$ modes are infrared active on the \textit{ab}-plane and the $A_2$ modes are silent. Early Raman data \cite{Haumont} found at least 11 modes at room temperature. Kamba \textit{et al.} \cite{Kamba} proposed to fit the infrared data on ceramics using 13 Lorentz oscillators. Although these results found the proper number of modes, they cannot clarify the symmetry of the phonons. Some extra insight on the phonon assignment was obtained on more recent Raman data \cite{Fukumura, Cazayous} but some inconsistencies in phonon assignment remain. In addition, Haumont \textit{et al.}\cite{Haumont} argued that because the frequency of the lowest mode did not vanish smoothly at the Curie temperature, the ferroelectric transition in BiFeO$_3$ would not be soft-mode driven. This picture was later revised to propose that the incomplete phonon softening could rather be a sign of a first order transition \cite{Kamba, Hermet}. However, both scenarios remain speculative.

In this paper, we show the polarized infrared response from 5 K to room temperature of a BiFeO$_3$ single crystal. Our crystal geometry allows us to unambiguously determine the \textit{ab}-plane $E$ phonons and to infer the frequencies of the \textit{c}-axis $A_1$ phonons. Moreover, we show that the temperature variation of the dc dielectric function is completely governed by the softening of the lowest frequency mode giving a strong evidence for a soft-mode driven transition.


High quality and purity BiFeO$_3$ crystals were obtained by flux growth \cite{Lebeugle, Lebeugle2007b}. We measured a thin BiFeO$_3$ plate measuring  $0.8 \times 0.8 \times 0.01$ mm$^3$. Near normal incidence infrared reflectivity was recorded from 25 to 4000\icm\ in a Bruker IFS66v interferometer. Below 650 \icm\ we utilized a Ge coated Mylar beamsplitter and a 4 K Si bolometer. Above 500 \icm\ a Ge:KBr beamsplitter and a MCT detector were used. In the overlapping region, the spectra agreed within 1 \%. The sample was mounted on the cold finger of an ARS Helitran open cycle cryostat equipped with polyethylene and KRS-5 windows. The spectra were collected on the as-grown $0.8 \times 0.8$ mm$^2$ flat face of the sample. To obtain the infrared reflectivity on such a small sample, an overfill technique with a \textit{in-situ} gold evaporation was performed \cite{Homes}. The measured surface, corresponding to the $(012)_{\rm hex}$ plane, is reminiscent of the high temperature cubic phase [$(010)_{\rm cub}$]. X-ray analysis on similar crystals, showed that the axis perpendicular to the sample flat surface makes an angle of $\sim 55^\circ$ with the \textit{c}-axis \cite{Lebeugle}. Using polarized light we could isolate the contribution from the \textit{ab}-plane. Our crystal geometry does not allow to isolate the \textit{c}-axis response which is heavily contaminated by the \textit{ab}-plane phonons. 


\begin{figure}
  \begin{center}
    \includegraphics[width=7.2cm]{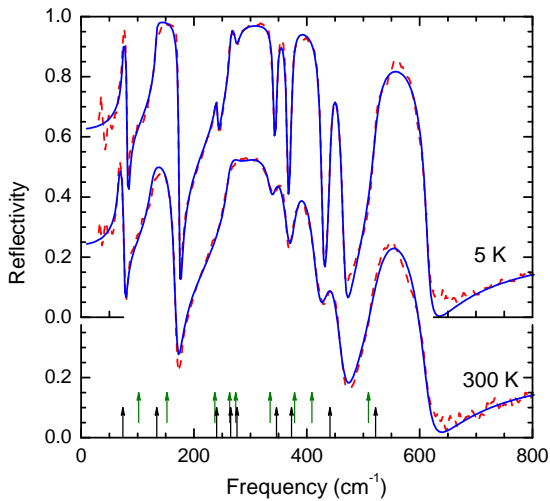}
  \end{center}
\caption{(Color online) In-plane (electric field of light $\mathbf{E} \perp \hat{z}$) infrared reflectivity of a BiFeO$_3$ single crystal at 5 K and 300 K. The dashed lines are the data and the solid lines are fits using a dielectric function in the form of Eq. \ref{eqDielFunc}. The lower arrows indicate the TO frequencies used in our fits. The upper arrows are the frequencies calculated in Ref. \onlinecite{Hermet}.}
\label{fig1}
\end{figure}

Figure \ref{fig1} shows the infrared reflectivity at 5 K and at room temperature for the electric field polarization lying in the \textit{ab}-plane. The dashed lines are the data and the solid lines are fits using a dielectric function of the form
\begin{equation}
\varepsilon\left(\omega\right) = \varepsilon_\infty + \sum_{j=1}^n \frac{\Delta\varepsilon_j \Omega_{TOj}}{\Omega_{TOj}^2 - \omega^2 -i \gamma_j \omega},
\label{eqDielFunc}
\end{equation}
where $\Omega_{TOj}$, $\gamma_j$ and $\Delta \varepsilon_j$ are, respectively, the transverse optical (TO) frequency, damping and oscillator strength for the $j$-th mode. $\varepsilon_\infty$ is the permittivity due to high frequency electronic excitations. The near normal incidence reflectivity is obtained from the Fresnel formula $R = \left|(\sqrt{\varepsilon}-1)/(\sqrt{\varepsilon}+1)\right|^2$. The fitting parameters for these two temperatures are shown in Table \ref{tab1}.

\begin{table}[htb]
  \caption{Fitting parameters used in Eq. \ref{eqDielFunc} to describe the in-plane ($\mathbf{E} \perp \hat{z}$) infrared reflectivity of a BiFeO$_3$ single crystal. The high frequency dielectric function ($\varepsilon_\infty$) was kept constant at 9.04 for all temperatures. Frequencies and dampings are in\icm\ and the oscillator strengths are dimensionless.}
  \begin{ruledtabular}
  \begin{tabular}{c......}
		\multicolumn{1}{c}{} & \multicolumn{1}{c}{} & \multicolumn{1}{c}{300 K} & \multicolumn{1}{c}{} & \multicolumn{1}{c}{} & \multicolumn{1}{c}{5 K} & \multicolumn{1}{c}{} \\
    \multicolumn{1}{c}{Mode} & \multicolumn{1}{c}{$\Omega_{TO}$}  & \multicolumn{1}{c}{$\gamma$} &\multicolumn{1}{c}{$\Delta \varepsilon$} &\multicolumn{1}{c}{$\Omega_{TO}$} &\multicolumn{1}{c}{$\gamma$} &\multicolumn{1}{c}{$\Delta \varepsilon$} \\
    \hline
    E(1) &  67 &  3.8 & 25.0   &  74 &  2.6 & 17.9 \\
    E(2) & 126 & 11.0 & 29.3   & 134 &  1.9 & 27.4 \\
    E(3) & 240 &  7.0 &  0.087 & 240 &  5.6 &  1.9 \\
    E(4) & 262 &  9.1 & 14.8   & 265 &  2.8 & 12.0 \\
    E(5) & 274 & 33.5 &  2.45  & 276 & 13.1 &  3.4 \\
    E(6) & 340 & 17.4 &  0.27  & 346 &  6.0 &  0.359 \\
    E(7) & 375 & 21.6 &  0.475 & 373 &  5.8 &  0.522 \\
    E(8) & 433 & 33.8 &  0.301 & 441 &  8.2 &  0.224 \\
    E(9) & 521 & 41.3 &  1.14  & 522 & 17.2 &  1.07 \\
  \end{tabular}
  \end{ruledtabular}
  \label{tab1}
\end{table}

The 5 K spectrum in Fig. \ref{fig1} clearly shows the 9 $E$ modes predicted by group theory. At room temperature, two of these phonons [$E(3)$ and $E(5)$] are strongly damped but keeping them improves the fit quality. Hermet \textit{et al.} \cite{Hermet} calculated the phonon frequencies expected for BiFeO$_3$ using density functional theory. The lower arrows (black) at the bottom of Fig. \ref{fig1} show the TO frequencies obtained from our fit at 5 K. The upper arrows (green) indicate the values calculated in Ref. \onlinecite{Hermet}. For most modes, the agreement is excellent (better than 5 \%). Modes $E(1)$ and $E(2)$, however, are found at frequencies 20--30\% smaller than the calculated values. Conversingly, mode $E(8)$ is found at an energy 8 \% higher than the theoretical value. Compared to the data reported by Kamba \textit{et al.} in a ceramic material \cite{Kamba}, our obtained frequencies are slightly upshifted, but the most important result concerns the modes assignment, once we only agree in the case of five modes, E(1), E(4), E(6), E(7) and E(9). The use of polarized light allowed us to correctly assign the other modes.

\begin{figure}
  \begin{center}
    \includegraphics[width=7.2cm]{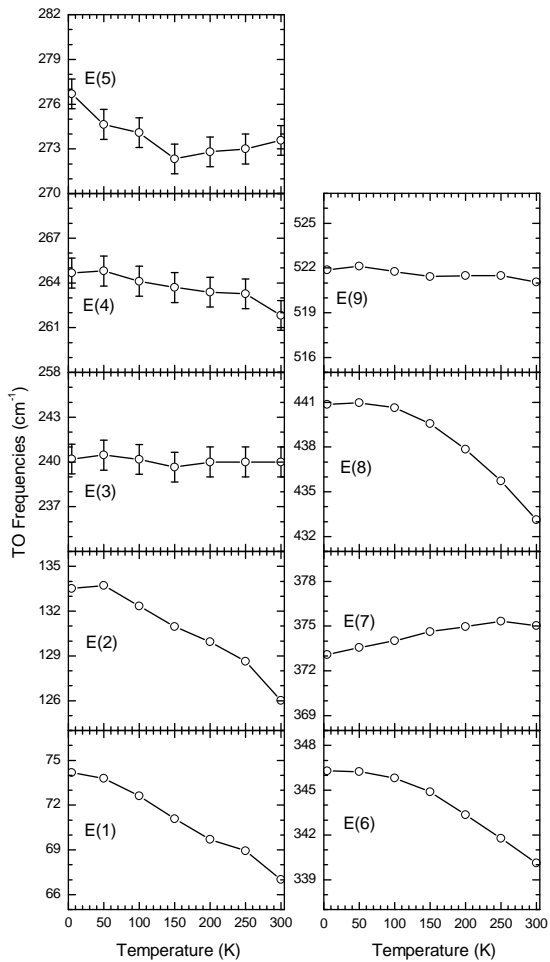}
  \end{center}
\caption{Temperature dependence of the in-plane ($\mathbf{E} \perp \hat{z}$) transverse optical frequencies for a BiFeO$_3$ single crystal. Due to its very small oscillator strength, the parameters for mode $E(3)$ have been kept constant above 200 K. Unless indicated otherwise, the values uncertainties are within the symbols size.}
\label{fig2}
\end{figure}

Figure \ref{fig2} shows the temperature evolution of the TO frequencies for the 9 $E$ phonon modes. The $E(1)$ mode softens strongly from 5 K to room temperature. High temperature infrared \cite{Kamba} and Raman \cite{Haumont, Fukumura} measurements indicate that this mode continues to soften but instead of smoothly decreasing to zero frequency it drops rather abruptly at the transition. Modes $E(2)$, $E(6)$ and $E(8)$ show the same behavior as mode $E(1)$. Interestingly, mode $E(7)$ \textit{hardens} with increasing temperature. The frequencies of the remaining modes have little or no temperature dependence. According to the calculated phonon frequencies \cite{Hermet}, $E(1)$ and $E(2)$ are related to vibrations involving the Bi atoms. $E(6)$ and $E(8)$ should be dominated by oxygen motion. Because of their softening, it is reasonable to think that these modes also involve motion of Bi atoms. Concerning E(7) --- also dominated by oxygen motion --- its hardening with increasing temperature shows that this mode has a particular dependence on the structural evolution, \textit{i.e.}, it is more sensitive to the decreasing rhombohedral elastic distortion than to the unit cell volume expansion, upon heating.

\begin{figure}
  \begin{center}
    \includegraphics[width=6cm]{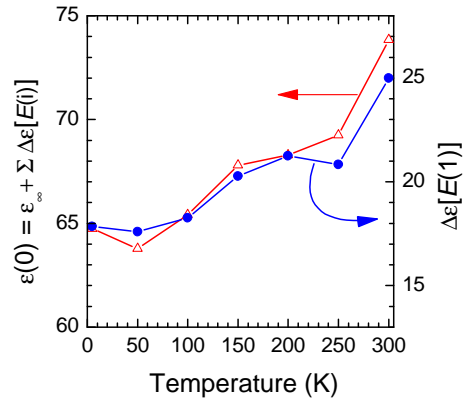}
  \end{center}
\caption{(Color online) The open triangles are the temperature dependence of the dielectric constant calculated using Eq. \ref{eqDielConst} (left axis). The solid circles depict the thermal evolution of the oscillator strength for the $E(1)$ mode (right axis). As both axis have the same range for $\varepsilon$, one can see that the temperature dependence of the dielectric constant comes mostly from the softening (hence oscillator strength increase) of the $E(1)$ phonon.}
\label{fig3}
\end{figure}

More insight on the $E(1)$ mode can be obtained from an analysis of its oscillator strength. One can calculate the dielectric constant from the optical contributions as
\begin{equation}
\varepsilon(0) = \varepsilon_\infty + \sum_{j=1}^n \Delta\varepsilon_j.
\label{eqDielConst}
\end{equation}
This calculation, for our results, is shown as open triangles in Fig. \ref{fig3}. Our values are twice the values obtained by Kamba \textit{et al.} \cite{Kamba}. This is a consequence of the higher reflectivity and absence of microstructure (porosity, grain boundaries) in our sample. The values in Fig. \ref{fig3} are a bit smaller (10 \%) than those obtained by dielectric measurements at 1 GHz on ceramics solid solutions of BiFeO$_3$ with Pb(Ti,Zr)O$_3$ \cite{Smith}. We also plotted in this figure the temperature dependence of the $E(1)$ mode oscillator strength (solid circles). It becomes clear that the temperature increase in $\varepsilon(0)$ is explained by the softening of the $E(1)$ mode alone. Indeed the \textit{f}-sum rule states that the sum of plasma frequencies squared ($\sum_j \Delta \varepsilon_j \Omega_{TOj}^2$) should be a constant independent of temperature \cite{Wooten}. An even more stringent condition is that, in the absence of phonon coupling, each individual plasma frequency should remain constant. Indeed, within $\pm 5 \%$, the plasma frequency for the mode $E(1)$ respects this condition. Therefore, the softening of the $E(1)$ phonon induces an increase in its oscillator strength which can fully account for the increase in the dielectric constant. This is a strong indication that this phonon participates actively as the soft mode driving the ferroelectric transition. As this mode dies sharply at the Curie temperature instead of vanishing smoothly \cite{Haumont,Kamba}, it was first considered that the transition was not soft-mode driven \cite{Haumont}. This picture was later revised to propose that the transition was displacive and \textit{did} have a soft mode which incomplete softening could be a sign of a first order transtion \cite{Kamba, Hermet}. Figure 3, gives a strong support for this latter proposition. 

An important point arises when we look at the theoretically obtained oscillator strengths and our experimentally determined values. Hermet \textit{et al.} \cite{Hermet} calculated the phonons TO and longitudinal optical (LO) frequencies and found $\Omega_{TO1} = 102\icm$ and $\Omega_{LO1} = 104\icm$ for mode $E(1)$ and $\Omega_{TO1} = 152\icm$ and $\Omega_{LO1} = 175\icm$ for mode $E(2)$. We can use the TO-LO splitting to calculate each mode oscillator strength \cite{Gervais}. The TO-LO splittings calculated by Hermet \textit{et al.}, yields theoretical oscillator strengths  of $\Delta\varepsilon_1 = 2.1$ and $\Delta\varepsilon_2 = 12.3$. This is strikingly different from the values we determined experimentally (see Table \ref{tab1}). We find that the oscillator strengths for the $E(1)$ and $E(2)$ modes are of the same order of magnitude and, more important, the oscillator strenght of mode $E(1)$ is large enough (by a factor 10 with respect to the calculation) to drive the changes in the dielectric constant. Our results, opposite to Hermet \textit{et al.} and in agreement with Kamba \textit{et al.}, show that the $E(1)$ mode is the soft mode in BiFeO$_3$.

Although we cannot measure directly the \textit{c}-axis response we can choose a polarization direction in our measurement crystal face that minimizes the \textit{ab}-plane contribution. We can thus infer the phonon locations along the \textit{c}-axis. 

\begin{figure}
  \begin{center}
    \includegraphics[width=7.2cm]{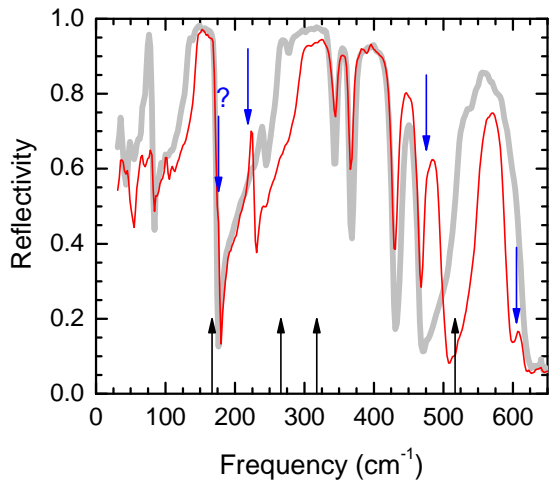}
  \end{center}
\caption{(Color online) The thinner line is the in-plane ($E \perp \hat{z}$) reflectivity. The thicker line is the combined response. The upward pointing arrows are the TO frequencies calculated in Ref. \onlinecite{Hermet}. The down pointing arrows are our proposition for the $E \parallel c$ phonons.}
\label{fig4}
\end{figure}

In Fig. \ref{fig4} we compare the \textit{ab}-plane response (thick gray line) to the mixed spectrum (thin red line) obtained at 5 K, temperature at which the phonons are sharp and well defined. There are three phonon modes that are very clearly present in the mixed response and absent from the \textit{ab}-plane spectrum. They are indicated by the down pointing arrows. The forth mode is very likely appearing as a broader shoulder just above the $E(2)$ mode ($\sim 180\icm$) indicated by the arrow with a question mark. We assign these features to the 4 $A_1$ modes. The upward pointing arrows at the bottom of the figure indicate the frequencies for the $A_1$ modes calculated from first principles \cite{Hermet}. The first two $A_1$ phonons found in our spectra (180\icm\ and 215\icm) agree relatively well with the predicted values. However, the phonon expected around 320\icm\ is found closer to 475\icm. The phonon expected at 520\icm\ appears above 600\icm. The disagreement for the latter phonon can be understood in terms of the strong $E(9)$ phonon contamination in our mixed response spectra. Because of this large $E(9)$ mode, the $A_1(4)$ mode may appear as a dip at its LO frequency, rather than a peak close the TO energy. The calculated LO frequency for the $A_1(4)$ mode is $\sim 535\icm$, closer to our finding but still a poorer agreement than the first two modes. We also remark that the assignment of the $A_1(3)$ and $A_1(4)$ modes differ substantially from previous propositions \cite{Haumont,Kamba}, although $A_1(3)$ agrees with recent Raman results \cite{Cazayous}.


We measured the low temperature polarized infrared response of a BiFeO$_3$ single crystal. We determined clearly all the \textit{ab}-plane phonon parameters. Our findings agree very well with calculated phonon frequencies and solve the phonon symmetry assignment issues. The \textit{c}-axis phonons could not be completely determined from our crystal geometry but their frequencies are inferred from a mixed polarization measurement. We show that the softening of the $E(1)$ phonon explains completely the temperature dependence of the dielectric constant. Our results are a clear evidence that the $E(1)$ phonon is undoubtedly the soft mode driving the ferroelectric transition.

\begin{acknowledgments}
This work was partially supported by the MIIAT proposal ``Mat\'eriaux \`a Propri\'et\'es Remarquables''.
RLM acknowledges support from the Brazilian agencies CNPq and FAPEMIG. During this work RLM was an invited professor at ESPCI on a ``Chaire Joliot'' appointment.
\end{acknowledgments}


\end{document}